\documentclass[conference]{IEEEtran}
\usepackage{cite}
\usepackage{amsmath,amssymb,amsfonts}
\usepackage{algorithm}
\usepackage{algorithmic}
\usepackage{graphicx}
\usepackage{textcomp}
\usepackage{xcolor}
\def\BibTeX{{\rm B\kern-.05em{\sc i\kern-.025em b}\kern-.08em
    T\kern-.1667em\lower.7ex\hbox{E}\kern-.125emX}}

\usepackage[hidelinks]{hyperref}
\usepackage{diagbox}
\usepackage[table]{xcolor}
\usepackage{multirow}
\usepackage{tikz}
\usetikzlibrary{quantikz2}
\usepackage{braket}
\usepackage{comment}

\begin{document}

\title{QUACOD: Quantum Optimization via Coordinate Descent for Scalable Drone Scheduling}

\author{
    Van-Quang-Huy Nguyen$^{1, 4}$, Hoang-Quan Nguyen$^{1, 4}$,
    Samee U. Khan$^2$, Ilya Safro$^3$, Khoa Luu$^{1, 4}$ \\
    $^1$Department of Electrical Engineering and Computer Science, University of Arkansas, Fayetteville, AR, USA \\
    $^2$Department of Electrical and Computer Engineering, Kansas State University, Manhattan, KS, USA \\
    $^3$Department of Computer and Information Sciences, University of Delaware, Newark, DE, USA \\
    $^4$Quantum AI Lab, University of Arkansas, Fayetteville, AR, USA \\
    \small{\texttt{\{hvn001, hn016, khoaluu\}@uark.edu}} \quad
    \small{\texttt{sameekhan@ksu.edu}} \quad \small{\texttt{isafro@udel.edu}}
}

\maketitle

\begin{abstract}
Quantum computing has demonstrated its potential to solve various optimization problems, including drone scheduling, which is important not only for drone delivery but also for logistics in general. However, one of the main obstacles is that practical drone scheduling settings typically require quantum resources that current hardware cannot provide. Therefore, in this work, we introduce a new Quantum Optimization via Coordinate Descent (QUACOD) approach to address this problem under the constraint of a limited number of available qubits. By leveraging coordinate descent, QUACOD decomposes the original high-complexity problem into multiple subproblems, which are then solved using quantum optimization. In our experiments, QUACOD outperforms the state-of-the-art (SOTA) quantum-based drone scheduling method not only in optimized drone completion times but also in scalability, handling up to 5 times more drones and 35 times more routes. In addition, QUACOD demonstrates that hardware-efficient circuits are effective for optimization problems. Together, these contributions advance quantum computing toward practical applications in the noisy intermediate-scale quantum (NISQ) era.

\end{abstract}


\section{Introduction}


With the rapid development of modern logistics and delivery systems necessitating more effective and economical solutions, particularly in last-mile logistics, unmanned aerial vehicles (UAVs), or drones, are increasingly recognized as a promising technology. Logistics companies actively explore drone-based delivery to reduce operational costs and improve delivery efficiency, thereby motivating a growing body of research on drone fleet management and route optimization \cite{jazairy2025drones}. More specifically, the literature \cite{jazairy2025drones, raivi2023drone} identifies several core problems central to drone logistics, including optimal, energy-constrained route planning and scheduling. These tasks are closely related to well-known combinatorial problems, such as the Traveling Salesman Problem, which belong to the NP-hard complexity class \cite{applegate2011traveling} and exhibit exponential growth in the solution space with problem size. While classical approaches, namely exact methods \cite{roberti2021exact} and metaheuristics such as genetic algorithms and ant colony optimization \cite{shuaibu2025review}, have been widely applied, the limitations of optimality and scalability as fleet sizes and delivery networks grow remain significant challenges, requiring more powerful computational approaches.


To overcome the inherently classical bottleneck in combinatorial optimization and related drone problems, quantum computing has emerged as an alternative approach, with promising demonstrations across related fields such as machine learning \cite{nguyen2024qclusformer, nguyen2025diffusion, nguyen2025qmoe}. By leveraging quantum effects to naturally explore large solution spaces, proposed quantum algorithms \cite{farhi2014QAOA, kadowaki1998quantum} are analytically shown to offer a computational advantage in solving such problems. However, applying quantum computing to optimization in practice remains challenging \cite{abbas2024challenges}. Many existing results are still theoretical, with experiments serving primarily as proof of concept. One of the main difficulties is the sensitivity of current Noisy Intermediate-Scale Quantum (NISQ) \cite{preskill2018quantum} devices to noise and decoherence, which limits the number and quality of qubits and hinders large and complex computations. For instance, current quantum hardware from vendors such as IBM and D-Wave ranges from a few hundred to a few thousand qubits \cite{ibm_quantum_hardware, dwave2025performance}, far from the millions required for truly large-scale problems. Solving practical problems without demanding excessive quantum resources has therefore become a prominent research direction, particularly in the NISQ era.

\textbf{Our Contributions in this Work:}
First, we propose a novel Quantum Optimization via Coordinate Descent (QUACOD) to solve the drone scheduling problem, which involves assigning routes to drones and becomes challenging as the problem size grows. Coordinate descent obtains the optimal solution by decomposing the problem into smaller subproblems, then solved using quantum optimization. By combining these two approaches, QUACOD unlocks the potential to solve larger problems that current quantum hardware cannot handle directly. Second, through experiments, QUACOD outperforms the state-of-the-art (SOTA) quantum framework for drone delivery (QUADRO) proposed by Holliday et al. \cite{holliday2025} on all drone scheduling instances in terms of maximum drone completion time. Finally, QUACOD demonstrates strong performance on large-scale instances of an adapted dataset, solving problems with up to 50 drones and 1,000 routes, which are 5 times more drones and 35 times more routes than QUADRO \cite{holliday2025}, using limited classical simulation resources and hardware-efficient circuits with the potential to be deployed on NISQ-era quantum hardware. These experimental results demonstrate the scalability of QUACOD and show that classical optimization strategies, such as coordinate descent, can further enhance quantum optimization, bringing it closer to practical applications.

The remainder of this work is organized as follows.
Section~\ref{background} reviews the relevant background and related work.
Section~\ref{methodology} covers the problem formulation, the necessary transformations, and the proposed method QUACOD to find the final optimal solution.
Section~\ref{experiments} describes the experiments and discusses the results achieved by QUACOD.
Section~\ref{conclusion} concludes the paper and summarizes the key findings.

\section{Background}\label{background}

\subsection{Drone Scheduling}

Delivery problems are typically divided into two main classes: ground vehicles, such as trucks and cars \cite{toth2014vehicle, golden2008vehicle, holliday2024hybrid, holliday2026advanced}, and drones \cite{dorling2016vehicle}.
In both classes, routing plays a central role and has been extensively studied. However, drone delivery mathematically introduces additional challenges. Due to engine constraints, drone delivery capacity is limited, so delivery tasks must be divided into significantly more routes than with ground vehicles. Also, drones are subject to battery constraints, forcing them to recharge between consecutive deliveries. This gives rise to the drone scheduling problem, which aims to assign and schedule routes across drones in an optimal manner.

\begin{figure}[t]
\centering
\includegraphics[width=\columnwidth]{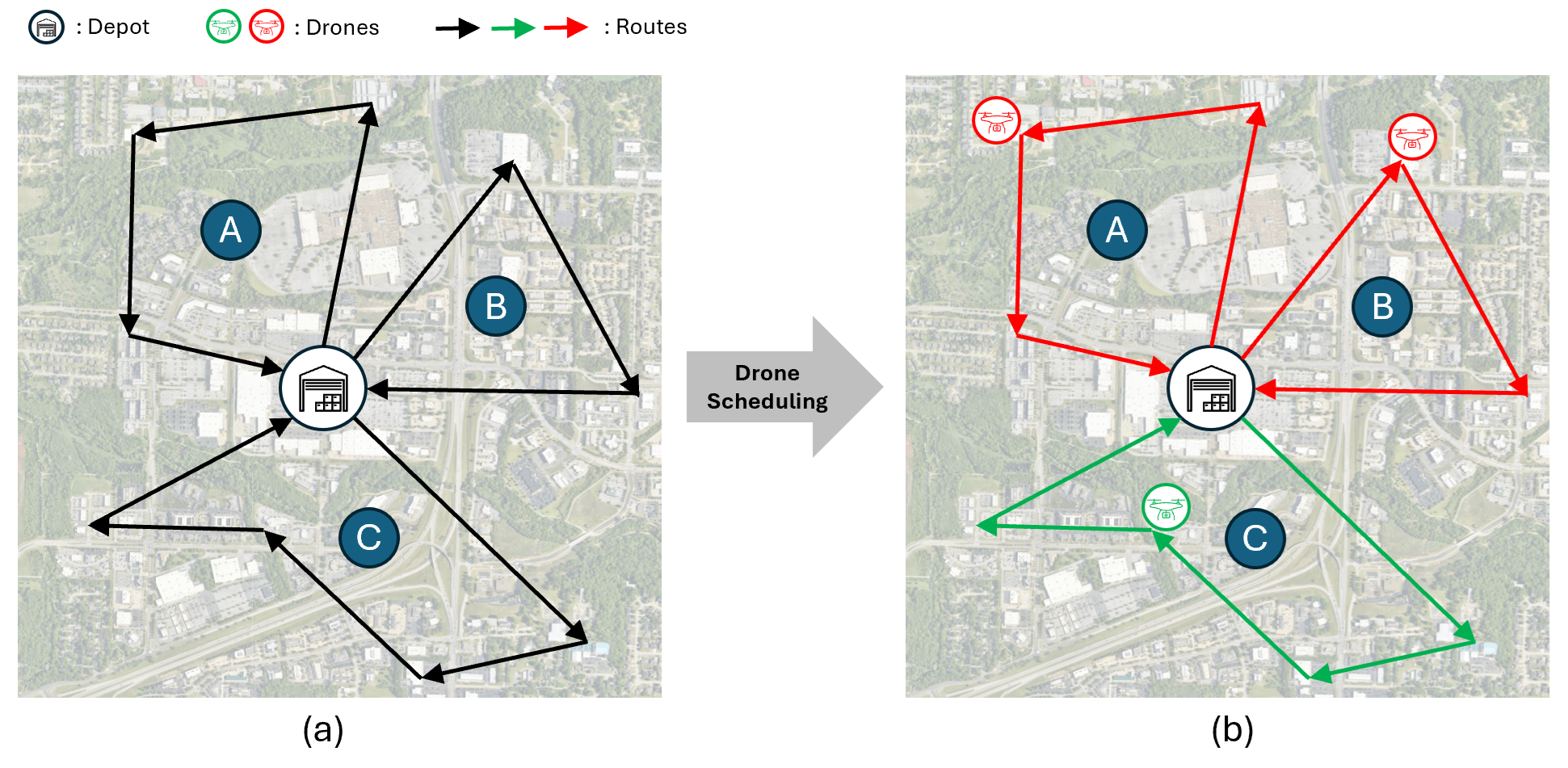}
\caption{Illustration of the drone scheduling problem with a charging time of 1.15 hours between routes. (a) Three predetermined routes, A, B, and C, have completion times of 3.4, 2.8, and 4.4 hours, respectively. (b) An example schedule assigns routes A and B to the red drone (completion time: $3.4 + 1.15 + 2.8 = 7.35$ hours) and route C to the green drone (completion time: 4.4 hours). The objective is to minimize the time until all drones have completed their routes, which equals 7.35 hours in this example.}
\label{drone_scheduling_fig}
\end{figure}

In this work, we focus on the drone scheduling problem, as illustrated in Fig.~\ref{drone_scheduling_fig}. The goal in assigning routes to drones is to minimize the time until all drones have completed their routes. Since scheduling does not involve routing, we assume that routes are computed in advance and their completion times are predetermined. In practice, drones are mass-produced and thus assumed to be identical, meaning route completion times apply uniformly across all drones. After completing a route, a drone must return to the depot to pick up packages for the next delivery and recharge its battery. Note that the recharging time is non-negligible and accounts for a significant portion of the total completion time of a drone. These characteristics distinguish the drone scheduling problem from traditional vehicle delivery problems.

\subsection{Quadratic Unconstrained Binary Optimization (QUBO)}

Quadratic unconstrained binary optimization (QUBO) is a class of quadratic models where the objective function contains terms of degree at most 2. However, QUBO has two notable properties. First, all variables are binary, and the problem is thus essentially discrete. Unlike continuous problems, where gradients can be exploited, binary problems are computationally hard.
In fact, some instances are even NP-hard (for example, max-cut \cite{karp2009reducibility}), making efficient optimization intractable in general. Second, QUBOs do not support constraints, which limit their direct applicability in practice and require several transformations before they can be used.
In short, QUBO can be expressed as in Eqn.~\eqref{qubo}.
\begin{equation}\label{qubo}
\underset{x_1,\ldots,x_n  \in \{0,1\}}{\text{minimize}} \quad \sum_{1 \leq i < j \leq n} a_{i,j} x_i x_j + \sum_{1 \leq i \leq n} b_i x_i
\end{equation}
where $a_{i, j}$ and $b_i$ are real coefficients.

A remarkable connection is that the QUBO objective function is equivalent to the Hamiltonian of Ising models \cite{lucas2014ising}. Consequently, minimizing the objective function corresponds to finding the ground state of the Ising model. Exploiting this equivalence opens the door to quantum approaches for solving optimization problems, such as variational quantum algorithms \cite{peruzzo2014variational, farhi2014QAOA} and quantum annealing \cite{kadowaki1998quantum}.

\subsection{Variational Quantum Algorithms}

\begin{figure}[t]
\centering
\includegraphics[width=\columnwidth]{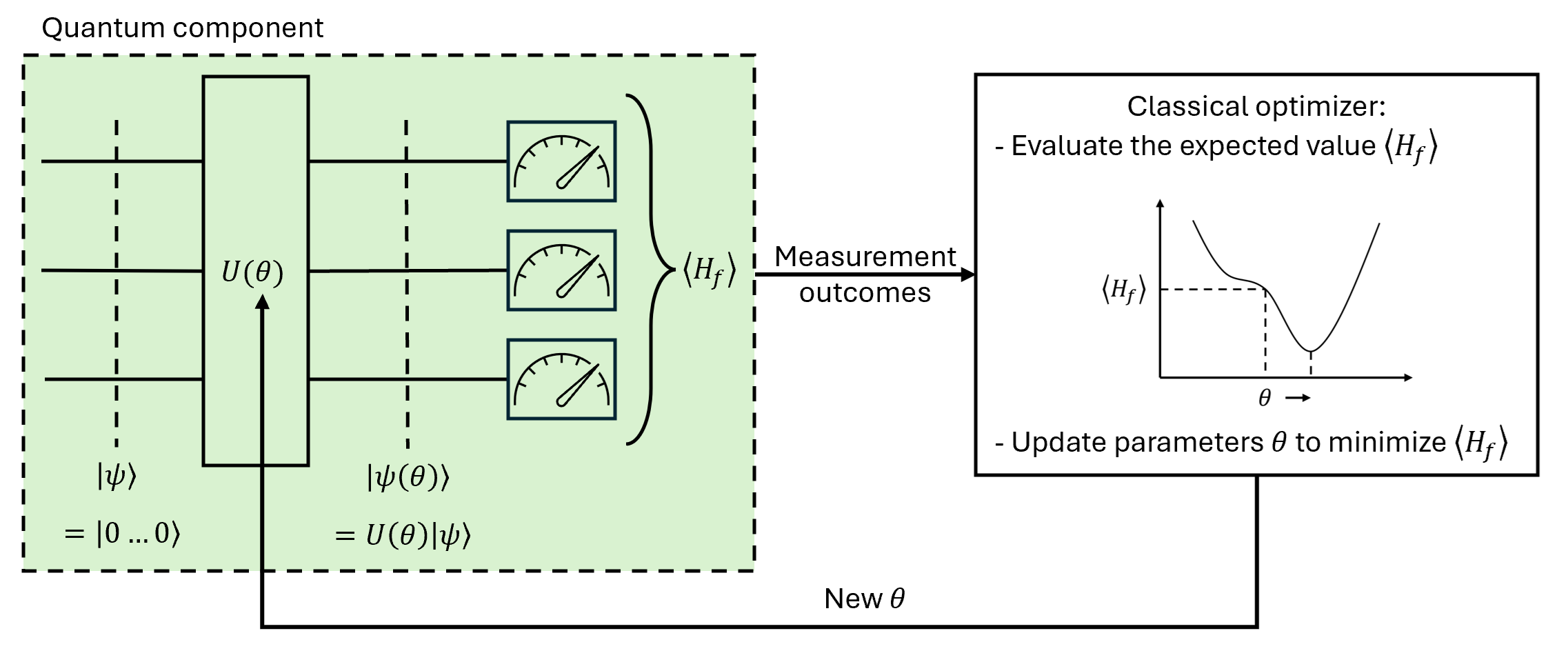}
\caption{The hybrid quantum-classical optimization loop of a variational quantum algorithm. The quantum component prepares a parameterized state $\ket{\psi(\theta)}$ and measures the expected value $\langle H_f \rangle$, which the classical optimizer uses to iteratively update $\theta$ until convergence.}
\label{variational_quantum_algorithms}
\end{figure}

Variational quantum algorithms \cite{cerezo2021variational} are hybrid optimizers that combine the strengths of both quantum and classical computational paradigms, as shown in Fig.~\ref{variational_quantum_algorithms}.
The quantum component consists of an ansatz, a variational quantum circuit $U(\theta)$, where $\theta$ are the circuit parameters. Initially, a quantum state $\ket{\psi} = \ket{0\ldots0}$ is prepared. After applying the circuit $U(\theta)$, the parameterized state $\ket{\psi(\theta)} = U(\theta)\ket{\psi}$ is obtained. The objective function $f$ is encoded as a Hamiltonian $H_f$, and its expectation value $\langle H_f \rangle = \bra{\psi(\theta)} H_f \ket{\psi(\theta)}$ is estimated from repeated measurements of the circuit. After evaluating the expectation value $\langle H_f \rangle$, a classical optimizer updates $\theta$ to minimize $\langle H_f \rangle$. This process is repeated until $\langle H_f \rangle$ converges to the ground state energy of $H_f$, corresponding to the minimum of $f$.
The optimal solution is then decoded from the measurement outcomes of the circuit at the optimized parameters.

The two most well-known instantiations of variational quantum algorithms are the Variational Quantum Eigensolver (VQE) \cite{peruzzo2014variational} and the Quantum Approximate Optimization Algorithm (QAOA) \cite{farhi2014QAOA}.
The key difference between the two lies in the ansatz construction.
The QAOA ansatz is derived directly from the objective function, making it well-suited for combinatorial optimization. In contrast, the VQE ansatz is chemically inspired, designed to estimate the ground-state energy of molecular systems. Although VQE is not specifically designed for combinatorial optimization, its convergence to the ground-state energy is equivalent to minimizing the objective function, making it applicable to this class of problems.

\subsection{Coordinate Descent Optimization}

Coordinate descent \cite{wright2015coordinate} is an iterative optimization algorithm that minimizes an objective function $f$ by updating one variable at a time while keeping the remaining variables fixed, as presented in Algorithm~\ref{coordinate_descent_alg}.
Unlike gradient descent, which updates all variables simultaneously at each iteration, coordinate descent cycles through each variable one by one.
At each iteration, the original problem reduces to a single-variable subproblem, keeping the computational cost substantially lower than that of full-gradient methods and making coordinate descent effective for large-scale, high-dimensional problems.

\begin{algorithm}
\caption{Coordinate Descent Algorithm\cite{wright2015coordinate}}
\label{coordinate_descent_alg}
\textbf{Input:} An objective function $f$ of $d$ variables $\mathbf{x}\!=\!(x_1,\ldots,x_d)$

\textbf{Output:} A vector $\mathbf{x}^{(k)} = (x^{(k)}_1, \ldots, x^{(k)}_d) = \underset{\mathbf{x}}{\arg\min}\ f(\mathbf{x})$

\begin{algorithmic}[1]
    \STATE Initialize $\mathbf{x}^{(0)} \gets (x^{(0)}_1, \ldots, x^{(0)}_d)$ and $k \gets 0$
    \REPEAT
        \STATE $k \gets k + 1$
        \FOR{$i = 1$ \TO $d$}\label{coordinate_descent_alg:for}
            \STATE $x^{(k)}_i \gets \underset{y}{\arg\min}\ f(x^{(k)}_1,\ldots,x^{(k)}_{i-1},y,x^{(k-1)}_{i+1},\ldots$ \\
            \hspace{7em} $\ldots,x^{(k-1)}_d)$\label{coordinate_descent_alg:main}
        \ENDFOR
    \UNTIL convergence is reached
    \RETURN $\mathbf{x}^{(k)}$
\end{algorithmic}
\end{algorithm}

Coordinate descent has demonstrated its convergence properties across various scenarios \cite{tseng2001convergence, luo1992convergence}. Several applications have been developed based on it, such as for support vector machines \cite{hsieh2008dual} in machine learning or generalized linear models \cite{friedman2010regularization} in computational statistics.
Various variants have been proposed to further improve the efficiency and applicability of coordinate descent, including randomized coordinate descent \cite{nesterov2012efficiency} and block coordinate descent \cite{tseng2001convergence}.
These works serve as the inspiration and foundation for QUACOD.

\subsection{Related Work}

Quantum studies and applications in drone delivery are relatively new.
QUADRO \cite{holliday2025} is considered the first to address problems related to this topic, proposing two main problems for drone delivery: (1) Energy-Constrained Capacitated Unmanned Aerial Vehicle Routing Problem (EC-CUAVRP) and (2) Unmanned Aerial Vehicle Scheduling Problem (UAVSP).
While quantum optimization has been successfully applied to EC-CUAVRP, it remains challenging for UAVSP.
In all experiments reported in \cite{holliday2025}, the maximum number of drones QUADRO could schedule did not exceed 11. Compared to the estimated minimum number of drones required for a hub in practice \cite{zieher2024drones}, this remains a significant limitation, motivating the need for a new method to address it, which is the goal of this work.

The scheduling challenge in UAVSP can be viewed as a large-scale optimization problem.
Such large-scale optimization problems can be broadly divided into two classes: those with a very large number of samples and those with a very large number of variables \cite{wright2022optimization}.
While the former is typically addressed by stochastic gradient descent, the latter is more suitable for coordinate descent.
Since UAVSP shares more properties with the latter class, coordinate descent is a natural approach.
It enables hybrid quantum-classical algorithms to solve larger problems with a limited number of qubits, which is the core idea behind QUACOD.

\section{Methodology}\label{methodology}

\subsection{Problem Formulation}

The drone scheduling problem involves assigning routes to drones optimally.
Given $n$ routes and $q$ drones in total, we assume $n > q$, since the solution is trivial if $n \leq q$.
The time for a drone to complete route $i$ is $r_i$, which may vary across routes. In contrast, all drones are identical.
For each route $i \in [n] = \{1,\ldots,n\}$ and each drone $j \in [q] = \{1,\ldots,q\}$, we define a binary variable $x_{i,j}$, where $x_{i,j} = 1$ if route $i$ is assigned to drone $j$, and $x_{i, j} = 0$ otherwise.
Then, the time for drone $j$ to complete all routes assigned to it is $\hat{T}_j$:
\begin{equation}
    \hat{T}_j = \sum_{i \in [n]} r_i x_{i,j}
\end{equation}
Since the operating time of drones is limited by battery constraints, they require recharging between consecutive routes. The recharging time is $c$.
We assume that there are no idle drones, meaning each drone is assigned at least one route. Then, the total recharging time for drone $j$ is $T^*_j$:
\begin{equation}
    T^*_j = c\left(\sum_{i \in [n]} x_{i,j} - 1\right)
\end{equation}
Since drones operate independently, the time required to complete all routes, $C_{max}$, is the maximum completion time $T_j = \hat{T}_j + T^*_j$ among all drones. In other words, this can be expressed as the constraint in Eqn.~\eqref{inequaility_constraints}.
\begin{equation}\label{inequaility_constraints}
\begin{aligned}
T_j &= \hat{T}_j + T^*_j \\
&= \sum_{i \in [n]} r_i x_{i,j} + c\left(\sum_{i \in [n]} x_{i,j} - 1\right)\leq C_{max}, \forall j \in [q]
\end{aligned}
\end{equation}
It is noted that each route is completed exactly once.
If route $i$ is assigned to drone $j$, no other drone is assigned to the same route, as per the constraint in Eqn.~\eqref{equaility_constraints}.
\begin{equation}\label{equaility_constraints}
    \sum_{j \in [q]} x_{i,j} = 1, \forall i \in [n]
\end{equation}
The objective of the drone scheduling problem is to minimize the time for all drones to complete all routes, i.e., minimizing $C_{max}$.
Following Holliday et al. \cite{holliday2025}, the problem is modeled as in Eqn.~\eqref{drone_scheduling_eqn}.
\begin{equation}\label{drone_scheduling_eqn}
\begin{aligned}
& \underset{\{x_{i,j}\}_{i \in [n], j \in [q]}}{\text{minimize}} & & C_{max} \\
& \text{subject to} & & x_{i,j} \in \{0,1\}, \hspace{2.35em} \forall i \in [n], j \in [q], \\
& & & \sum_{j \in [q]} x_{i,j} = 1, \hspace{2.15em} \forall i \in[n], \\
& & & \sum_{i \in [n]} r_i x_{i,j} + c\left(\sum_{i \in [n]} x_{i,j} - 1\right) \\
& & & \qquad \leq C_{max}, \qquad \forall j \in[q]. \\
\end{aligned}
\end{equation}

\subsection{QUBO Reformulation}

To apply hybrid quantum-classical optimization, the problems should be expressed in QUBO form.
However, the drone scheduling problem modeled in Eqn.~\eqref{drone_scheduling_eqn} does not satisfy this requirement, as it contains equality and inequality constraints that must be transformed before proceeding to the next steps.

The only equality constraints in the drone scheduling problem are those requiring each route to be assigned to exactly one drone, as expressed in Eqn.~\eqref{equaility_constraints}.
Following the strategy in \cite{lucas2014ising, verma2022penalty}, any solution satisfying Eqn.~\eqref{equaility_constraints} minimizes the following term:
\begin{equation}
    \sum_{i \in [n]}\left(\sum_{j \in [q]} x_{i, j} - 1\right)^2
\end{equation}
By introducing a hyperparameter $p$ as a penalty, we define a new objective function to minimize:
\begin{equation}\label{intermediate_obj_func}
    C_{max} + p \sum_{i \in [n]}\left(\sum_{j \in [q]} x_{i, j} - 1\right)^2
\end{equation}

\begin{figure}[t]
\centering
\includegraphics[width=\columnwidth]{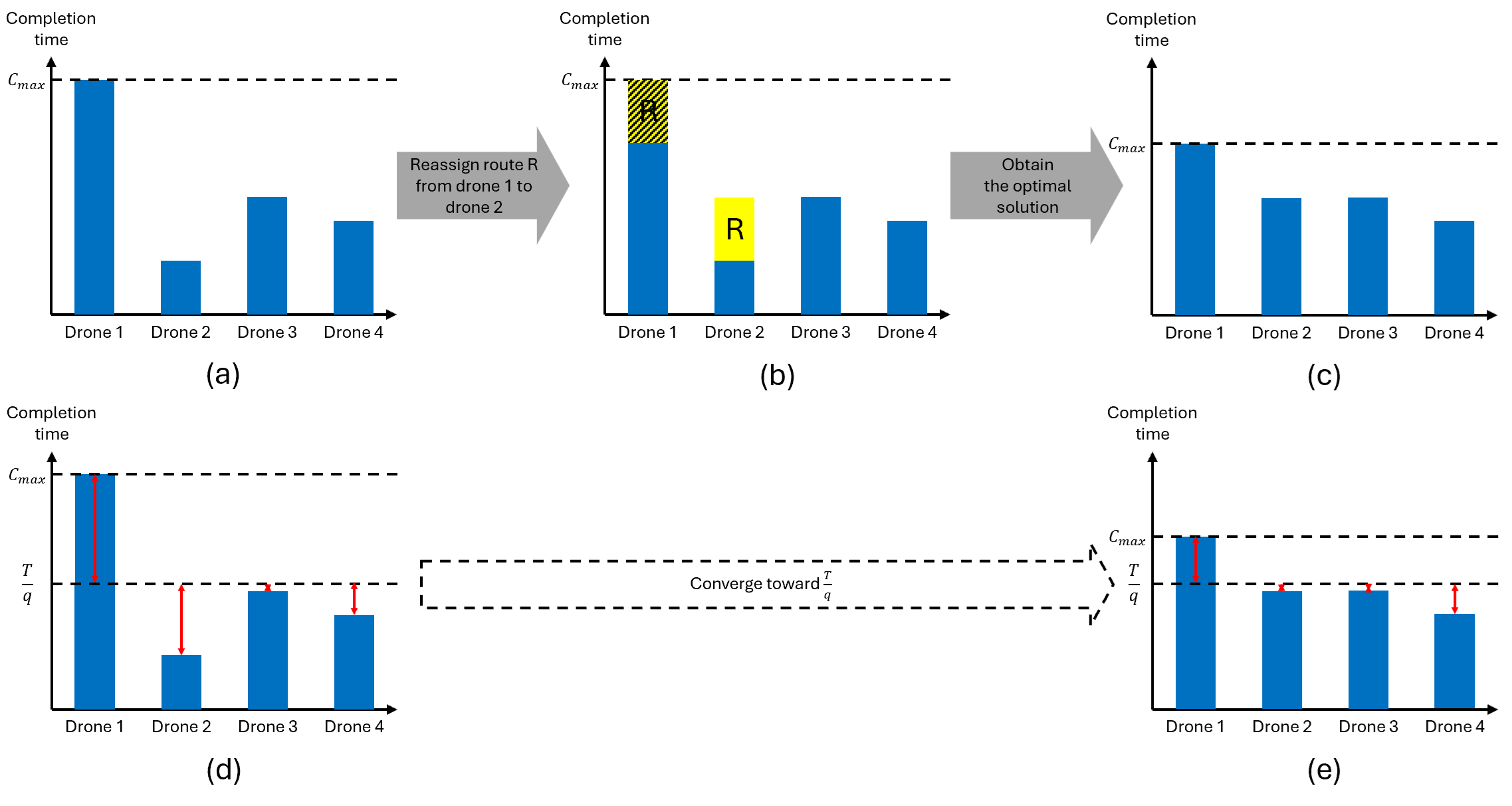}
\caption{Heuristic for transformation from $C_{max}$ minimization to convergence toward $T/q$. (a)-(c) Route assignment minimizes $C_{max}$ toward the optimal solution. (d)-(e) Minimizing the sum of squared distances $\sum_{j\in[q]}\left(T_j - \frac{T}{q}\right)^2$ instead of $C_{max}$ converges toward the same optimal solution.}
\label{heuristic_Cmax_to_Tq}
\end{figure}

Next, to eliminate the inequality constraints in Eqn.~\eqref{inequaility_constraints}, we reuse the heuristic proposed by Holliday et al. \cite{holliday2025}. More specifically, we find that the sum of all values $T_j$ is constant:
\begin{equation}
    T = \sum_{j \in [q]} T_j = \sum_{i \in [n]} r_i +c(n-q) = const
\end{equation}
As illustrated in Fig.~\ref{heuristic_Cmax_to_Tq}, minimizing $C_{max}$ forces all values $T_j$ to converge to the average $\frac{T}{q}$.
Motivated by this observation and following Holliday et al. \cite{holliday2025}, we instead minimize the sum of squared distances between each $T_j$ and $\frac{T}{q}$.
The objective function in Eqn.~\eqref{intermediate_obj_func} can thus be rewritten as:
\begin{equation}\label{final_obj_func}
\begin{split}
f\left(\{x_{i,j}\}\right) 
&= \sum_{j\in[q]}\left(T_j - \frac{T}{q}\right)^2 + p \sum_{i \in [n]}\left(\sum_{j \in [q]} x_{i, j} - 1\right)^2 \\
&= \sum_{j\in[q]}\left(\sum_{i \in [n]} r_i x_{i,j} + c\left(\sum_{i \in [n]} x_{i,j} - 1\right) - \frac{T}{q}\right)^2 \\
&\quad + p \sum_{i \in [n]}\left(\sum_{j \in [q]} x_{i, j} - 1\right)^2
\end{split}
\end{equation}
which is the objective function to be minimized in a QUBO form.

\subsection{Quantum Optimization via Coordinate Descent (QUACOD)}\label{QUACOD_section}

We propose QUACOD, a novel hybrid quantum-classical optimization algorithm based on coordinate descent \cite{wright2015coordinate}, to solve the drone scheduling problem. The objective function $f$ is then can be expressed as in Eqn.~\eqref{final_obj_func}, and the  Algorithm~\ref{QUACOD_alg} details the proposed algorithm implementation. To optimize $f$, one can use variational quantum algorithms, such as VQE \cite{peruzzo2014variational} or QAOA \cite{farhi2014QAOA}. However, these algorithms require the number of qubits to be at least equal to the number of variables, and since $f$ contains $nq$ binary variables in total, the number of required qubits becomes impractically large as the number of drones or routes grows.
QUACOD is designed to overcome this limitation, specifically as follows:

\begin{algorithm}
\caption{The QUACOD Algorithm}
\label{QUACOD_alg}
\textbf{Input:} $f$ in Eqn.~\eqref{final_obj_func} of $nq$ variables $\mathbf{x} = \{x_{i,j}\}_{i \in [n], j \in [q]}$

\textbf{Output:} $\mathbf{x}^{(k)} = \{x^{(k)}_{i,j}\}_{i \in [n], j \in [q]} = \underset{\mathbf{x}}{\arg\min}\ f(\mathbf{x})$

\begin{algorithmic}[1]
    \STATE Initialize $\{x^{(0)}_{i,j}\}_{i \in [n], j \in [q]}$ and $k \gets 0$
    \REPEAT
        \STATE $k \gets k + 1$
        \STATE Randomly select $N \subset [n]$ and $Q \subset [q]$ satisfying Eqns. \eqref{qubit_condition}, \eqref{exist_one_condition}, \eqref{exist_max_condition}, \eqref{exist_min_condition} \label{QUACOD_alg:selection}
        \STATE $\{x^{(k)}_{i,j}\}_{i \in [n] \setminus N, j \in [q] \setminus Q} \gets \{x^{(k-1)}_{i,j}\}_{i \in [n] \setminus N, j \in [q] \setminus Q}$ \label{QUACOD_alg:fixed}
        \STATE $\{x^{(k)}_{i,j}\}_{i \in N, j \in Q} \gets $ \\
        \hspace{3em} $\text{VQE}\left(\{x_{i,j}\}_{i \in N, j \in Q}; f, \{x^{(k)}_{i,j}\}_{i \in [n] \setminus N, j \in [q] \setminus Q}\right)$ \label{QUACOD_alg:optimized}
    \UNTIL convergence
    \RETURN $\mathbf{x}^{(k)}$
\end{algorithmic}
\end{algorithm}

\begin{itemize}
    \item Instead of iterating over variables sequentially (line~\ref{coordinate_descent_alg:for} of Algorithm~\ref{coordinate_descent_alg}), we select a subset of variables satisfying certain conditions (line~\ref{QUACOD_alg:selection} of Algorithm~\ref{QUACOD_alg}).
    \item Unselected variables are fixed at line~\ref{coordinate_descent_alg:main} of Algorithm~\ref{coordinate_descent_alg}, which is analogously done at line~\ref{QUACOD_alg:fixed} of Algorithm~\ref{QUACOD_alg}.
    \item Only the selected variables are optimized, as shown at line~\ref{coordinate_descent_alg:main} of Algorithm~\ref{coordinate_descent_alg} and line~\ref{QUACOD_alg:optimized} of Algorithm~\ref{QUACOD_alg}.
    \item The $\arg\min\ f$ function of Algorithm~\ref{coordinate_descent_alg} is realized through a hybrid quantum-classical optimization (line~\ref{QUACOD_alg:optimized} of Algorithm~\ref{QUACOD_alg}), which is feasible since the number of selected variables can be bounded.
\end{itemize}

As for selecting a subset of variables, not every subset is appropriate. Since we use variational quantum algorithms for optimization, the subset size must not exceed the number of qubits, $m$.
Next, to reflect the transformation as depicted in Fig.~\ref{heuristic_Cmax_to_Tq}, the subset of variables is formed from certain routes and drones. Given such a subset of routes $N$ and such a subset of drones $Q$, the subset of variables is ${\{x_{i,j}\}_{i \in N, j \in Q}}$. Furthermore, to ensure the algorithm yields non-trivial solutions at iteration $k$, for all $i \in N$, there exists one $j \in Q$ such that $x^{(k - 1)}_{i, j}= 1$. These conditions are illustrated by an example in Table~\ref{selecting_a_subset_of_variables_tab}, where there are $m=10$ qubits, $n=4$ routes, and $q=7$ drones. The selected route subset is $N = \{2,3\}$, the selected drone subset is $Q = \{2, 4, 5, 6\}$, and the variables to be optimized are represented as asterisks. If the value 1 were to appear in the fixed variables for route $i=2$ and route $i=3$, variables $x^{(k - 1)}_{2, j}$ and $x^{(k - 1)}_{3, j}$ for $j \in Q$ would be zero, which is trivial. This explains why no entry of 1 should appear in the fixed variables for route $i=2$ and route $i=3$. The number of variables is $|N \times Q| = 8$, which also satisfies the condition of not exceeding $m=10$.

\begin{table}[b]
    \centering
    \caption{Example of selecting a subset of variables, where asterisks represent the variables to be optimized and bold values indicate the current assignments $x^{(k-1)}_{i,j} = 1$.}
    \label{selecting_a_subset_of_variables_tab}
    \renewcommand{\arraystretch}{1.3}
    \begin{tabular}{|c|c|c|c|c|c|c|c|}
        \hline
        \diagbox{Route $i$}{Drone $j$} & 1 & 2 & 3 & 4 & 5 & 6 & 7 \\
        \hline
        1 & 0 & 0 & 0 & \textbf{1} & 0 & 0 & 0 \\
        \hline
        2 & 0 & \cellcolor{gray!20}$*$ & 0 & \cellcolor{gray!20}$*$ & \cellcolor{gray!20}$*$ & \cellcolor{gray!20}$*$ & 0 \\
        \hline
        3 & 0 & \cellcolor{gray!20}$*$ & 0 & \cellcolor{gray!20}$*$ & \cellcolor{gray!20}$*$ & \cellcolor{gray!20}$*$ & 0 \\
        \hline
        4 & \textbf{1} & 0 & 0 & 0 & 0 & 0 & 0 \\
        \hline
    \end{tabular}
\end{table}

Since QUACOD relies on random selections of route subset $N$ and drone subset $Q$, it may fail to yield any improvements at certain iterations when the completion times of drones in $Q$ are nearly identical.
To mitigate this and accelerate convergence, we add the condition that $Q$ must include the drones with the highest and lowest completion times.
The drone with the highest completion time is the current bottleneck and thus a natural candidate for rescheduling.
The drone with the lowest completion time, on the other hand, is the most under-loaded and can potentially absorb routes from other drones, facilitating a more balanced redistribution. In summary, given a quantum system of $m$ qubits, the subsets $N \subset [n]$ and $Q \subset [q]$ selected randomly at iteration $k$ must satisfy the following conditions:
\begin{equation}\label{qubit_condition}
    |N \times Q| \leq m
\end{equation}
\begin{equation}\label{exist_one_condition}
    \forall i \in N, \exists j \in Q, x^{(k-1)}_{i, j} = 1
\end{equation}
\begin{equation}\label{exist_max_condition}
    \underset{j}{\arg\max}\ T^{(k-1)}_j \in Q
\end{equation}
\begin{equation}\label{exist_min_condition}
    \underset{j}{\arg\min}\ T^{(k-1)}_j \in Q
\end{equation}
The subset of variables to be updated is then ${\{x_{i,j}\}_{i \in N, j \in Q}}$.

\begin{figure}[t]
\centering
\begin{quantikz}
    \lstick{$\ket{0}$} & \gate{R_y(\theta_{0,1})} & \ctrl{1} & \qw & \qw & \qw & \gate{R_y(\theta_{1,1})} & \qw \\
    \lstick{$\ket{0}$} & \gate{R_y(\theta_{0,2})} & \control{} & \ctrl{1} & \qw & \qw & \gate{R_y(\theta_{1,2})} & \qw \\
    \lstick{$\ket{0}$} & \gate{R_y(\theta_{0,3})} & \qw & \control{} & \ctrl{1} & \qw & \gate{R_y(\theta_{1,3})} & \qw \\
    \lstick{$\vdots$} & \setwiretype{n}\vdots & \setwiretype{n} & \setwiretype{n} & \setwiretype{n}\vdots & \setwiretype{n}\vdots & \setwiretype{n}\vdots & \setwiretype{n} \\
    \lstick{$\ket{0}$} & \gate{R_y(\theta_{0,m})} & \qw & \qw & \qw & \control{} & \gate{R_y(\theta_{1,m})} & \qw
\end{quantikz}
\caption{A single-layer hardware-efficient ansatz for an $m$-qubit register, comprising two $R_y$ rotation blocks with a linear $CZ$ entanglement chain on consecutive qubit pairs in between. Each $R_y$ gate has its own trainable parameter, giving $2m$ parameters $\theta_{0,1}, \ldots, \theta_{0,m}, \theta_{1,1}, \ldots, \theta_{1,m}$ in total.}
\label{ansatz_fig}
\end{figure}
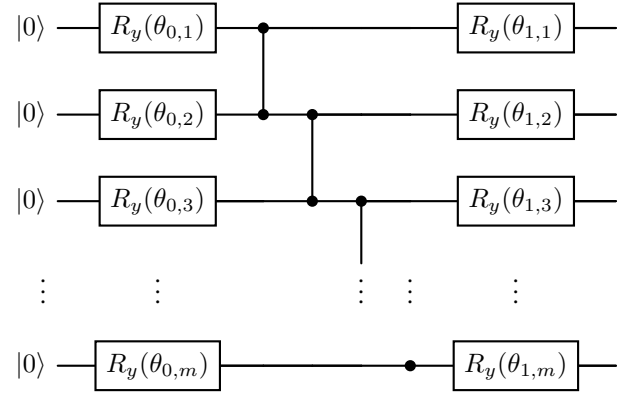

\begin{figure*}[t]
\centering
\includegraphics[width=\textwidth]{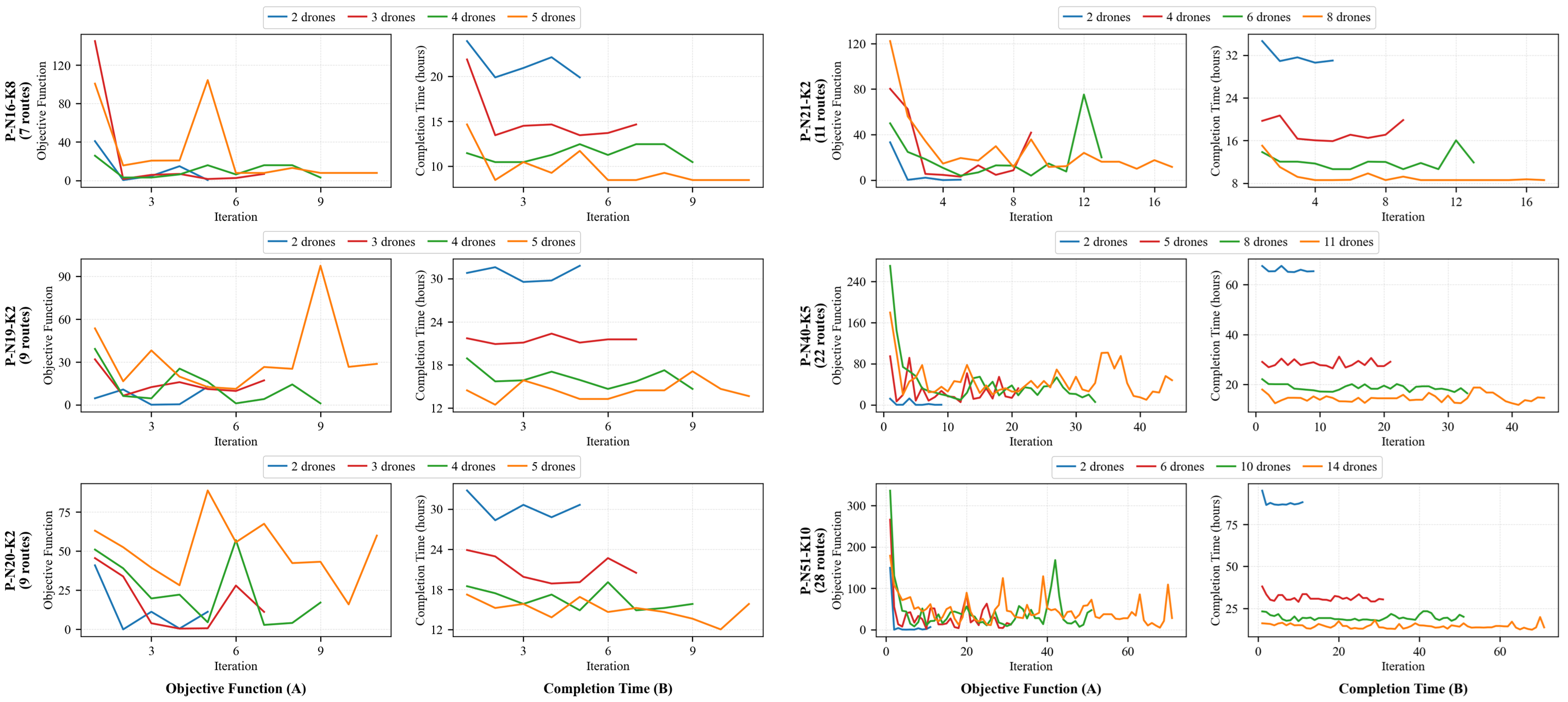}
\caption{Convergence process of QUACOD on the Augerat dataset \cite{augerat1995computational} adapted for the drone delivery problems \cite{uchoa2017new}. Column (A) shows the objective function value per iteration;  Column (B) shows the time (in hours) for all drones to complete
all routes per iteration.}
\label{convergence_plots_on_QUADRO_dataset}
\end{figure*}

As for selecting a hybrid quantum-classical algorithm to optimize $f$ in Eqn.~\eqref{final_obj_func}, there are several approaches, including variational quantum algorithms \cite{peruzzo2014variational, farhi2014QAOA} and quantum annealing \cite{kadowaki1998quantum}.
Although quantum annealing is a promising approach, no known libraries support its simulation on both CPUs and GPUs, limiting its simulation capability compared to gate-based quantum computing libraries.
Thus, we focus on variational quantum algorithms, of which VQE \cite{peruzzo2014variational} and QAOA \cite{farhi2014QAOA} are the most notable.
However, \cite{farhi2020quantum, bravyi2020obstacles} demonstrated that QAOA may require a large number of layers to produce quality solutions, which significantly increases the computational cost. Moreover, each layer of the QAOA ansatz is constructed from products of unitaries corresponding to the cost and mixer Hamiltonians, where the cost Hamiltonian is derived from the objective function. As the number of terms in the objective function grows, as in $f$ in Eqn.~\eqref{final_obj_func}, the cost unitary requires increasingly more single-qubit $Z$ gates and two-qubit $ZZ$ gates, further compounding the computational cost. In contrast, VQE offers simpler and more flexible ansatzes that are independent of the cost Hamiltonian. Since the VQE ansatz has more parameters than the QAOA ansatz, even a single-layer VQE can produce high-quality solutions. Its effectiveness on practical problems has been demonstrated in \cite{oh2019solving, liu2022layer}. Therefore, in this work, we adopt the VQE algorithm.

Among the VQE ansatz families, we choose the hardware-efficient one proposed in \cite{kandala2017hardware}. This family not only allows faster computation, but is also compatible with near-term quantum devices, consistent with our goal of developing algorithms for the NISQ era.
Fig.~\ref{ansatz_fig} shows a single-layer ansatz of the family used in QUACOD, comprising two $R_y$ rotation blocks with a $CZ$ linear entanglement block in between.
Each $R_y$ gate has its own trainable parameter, so in the case of $m$ qubits, the ansatz has $2m$ trainable parameters $\theta_{0,1}, \ldots, \theta_{0,m}, \theta_{1,1}, \ldots, \theta_{1,m}$ in total, providing sufficient expressibility to find high-quality solutions even with such a single layer.
Note that $R_y$ and $CZ$ gates with a linear entanglement structure are not required, and can be replaced with any rotation and controlled gates, as well as other entanglement structures. Everything else of VQE remains the same as in Fig.~\ref{variational_quantum_algorithms}.

\section{Experiments and Results}\label{experiments}

\subsection{Experimental Setup}

We simulate a quantum environment on a classical computer to evaluate QUACOD. Among open-source quantum simulation frameworks, we select Qiskit \cite{javadi2024quantum} for this purpose. Qiskit supports a wide range of features, including numerous simulation methods that can simulate up to tens or even hundreds of qubits on both CPU and GPU, leveraging the cuQuantum \cite{bayraktar2023cuquantum} library. Furthermore, selecting Qiskit is consistent with QUADRO \cite{holliday2025}, which we use as our baseline for comparison.

Although there are no specific constraints on the hyperparameters of QUACOD, we fix them to ensure consistency across all experiments.
We set the number of qubits to $m = 20$, which is large enough for quantum simulation while keeping execution time manageable.
The penalty is set to $p = \sum_{i \in [n]} r_i + cn$ to balance between constraint satisfaction and solution optimality.
Finally, the number of VQE ansatz layers used in QUACOD is set to $1$ (as illustrated in Fig.~\ref{ansatz_fig}), which is sufficient to find the optimal solution.

All experiments are executed on a workstation running Ubuntu 24.04 LTS (Linux kernel 6.14.0). The hardware configuration consists of an AMD Ryzen 7 5800X processor (8 cores, 16 threads, 3.8 GHz base clock), an NVIDIA GeForce RTX 3060 GPU featuring 12 GB of GDDR6 VRAM, and 16 GB of RAM. The setup runs using Python 3.12.

\subsection{Comparison with State-of-the-Art Methods} \label{QUADRO}

\begin{table*}[t]
\caption{Completion time (in hours) of QUADRO \cite{holliday2025} and QUACOD on the Augerat dataset \cite{augerat1995computational} adapted for the drone delivery problems \cite{uchoa2017new}, where the first columns show the hybrid and pure quantum results from QUADRO, and the last column shows QUACOD's results (lower is better).}
\centering
\resizebox{\textwidth}{!}{
\begin{tabular}{|c|c|c|c|c|c|c|c|c|c|c|c|c|c|c|c|}
\hline
\multirow{3}{*}{\textbf{Drones}} & \multicolumn{3}{c|}{\textbf{P-N16-K8 (7 routes)}} & \multicolumn{3}{c|}{\textbf{P-N19-K2 (9 routes)}} & \multicolumn{3}{c|}{\textbf{P-N20-K2 (9 routes)}} & \multicolumn{2}{c|}{\textbf{P-N21-K2 (11 routes)}} & \multicolumn{2}{c|}{\textbf{P-N40-K5 (22 routes)}} & \multicolumn{2}{c|}{\textbf{P-N51-K10 (28 routes)}} \\
\cline{2-16}
& \multicolumn{2}{c|}{\textbf{QUADRO \cite{holliday2025}}} & \multirow{2}{*}{\textbf{QUACOD}} & \multicolumn{2}{c|}{\textbf{QUADRO \cite{holliday2025}}} & \multirow{2}{*}{\textbf{QUACOD}} & \multicolumn{2}{c|}{\textbf{QUADRO \cite{holliday2025}}} & \multirow{2}{*}{\textbf{QUACOD}} & \textbf{QUADRO \cite{holliday2025}} & \multirow{2}{*}{\textbf{QUACOD}} & \textbf{QUADRO \cite{holliday2025}} & \multirow{2}{*}{\textbf{QUACOD}} & \textbf{QUADRO \cite{holliday2025}} & \multirow{2}{*}{\textbf{QUACOD}} \\
\cline{2-3} \cline{5-6} \cline{8-9} \cline{11-11} \cline{13-13} \cline{15-15}
& \textbf{Hybrid} & \textbf{Quantum} & & \textbf{Hybrid} & \textbf{Quantum} & & \textbf{Hybrid} & \textbf{Quantum} & & \textbf{Hybrid} & & \textbf{Hybrid} & & \textbf{Hybrid} & \\
\hline
2 & 38.85 & 21.25 & \textbf{19.90} & 50.20 & 31.76 & \textbf{29.55} & 44.63 & 29.60 & \textbf{28.35} & 57.62 & \textbf{30.60} & 104.33 & \textbf{65.05} & 153.03 & \textbf{86.45} \\
\hline
3 & 18.23 & 16.95 & \textbf{13.45} & 25.13 & $-$ & \textbf{20.90} & 29.22 & $-$ & \textbf{18.90} & 34.08 & \textbf{21.05} & 62.48 & \textbf{43.60} & 94.12 & \textbf{57.70} \\
\hline
4 & 17.43 & $-$ & \textbf{10.45} & 28.87 & $-$ & \textbf{14.65} & 33.18 & $-$ & \textbf{14.90} & 27.43 & \textbf{15.90} & 73.88 & \textbf{32.90} & 86.62 & \textbf{43.95} \\
\hline
5 & 14.20 & $-$ & \textbf{8.45} & 18.35 & $-$ & \textbf{13.25} & 18.40 & $-$ & \textbf{12.05} & 26.87 & \textbf{12.05} & $-$ & \textbf{26.45} & $-$ & \textbf{34.40} \\
\hline
6 & 10.03 & $-$ & \textbf{8.20} & 17.70 & $-$ & \textbf{11.05} & 15.80 & $-$ & \textbf{10.45} & 18.72 & \textbf{10.65} & $-$ & \textbf{22.15} & $-$ & \textbf{28.95} \\
\hline
7 & \textbf{8.20} & $-$ & \textbf{8.20} & 12.72 & $-$ & \textbf{10.05} & 14.52 & $-$ & \textbf{8.80} & 17.58 & \textbf{9.25} & $-$ & \textbf{19.35} & $-$ & \textbf{25.05} \\
\hline
8 & $-$ & $-$ & $-$ & 11.63 & $-$ & \textbf{8.60} & 12.63 & $-$ & \textbf{8.80} & 16.50 & \textbf{8.60} & $-$ & \textbf{16.45} & $-$ & \textbf{21.75} \\
\hline
9 & $-$ & $-$ & $-$ & \textbf{8.60} & $-$ & \textbf{8.60} & \textbf{8.80} & $-$ & \textbf{8.80} & 19.62 & \textbf{8.60} & $-$ & \textbf{15.20} & $-$ & \textbf{19.35} \\
\hline
10 & $-$ & $-$ & $-$ & $-$ & $-$ & $-$ & $-$ & $-$ & $-$ & 10.08 & \textbf{8.60} & $-$ & \textbf{13.25} & $-$ & \textbf{17.70} \\
\hline
11 & $-$ & $-$ & $-$ & $-$ & $-$ & $-$ & $-$ & $-$ & $-$ & \textbf{8.60} & \textbf{8.60} & $-$ & \textbf{11.80} & $-$ & \textbf{16.10} \\
\hline
12 & $-$ & $-$ & $-$ & $-$ & $-$ & $-$ & $-$ & $-$ & $-$ & $-$ & $-$ & $-$ & \textbf{11.50} & $-$ & \textbf{14.50} \\
\hline
13 & $-$ & $-$ & $-$ & $-$ & $-$ & $-$ & $-$ & $-$ & $-$ & $-$ & $-$ & $-$ & \textbf{11.50} & $-$ & \textbf{13.45} \\
\hline
14 & $-$ & $-$ & $-$ & $-$ & $-$ & $-$ & $-$ & $-$ & $-$ & $-$ & $-$ & $-$ & \textbf{11.50} & $-$ & \textbf{12.45} \\
\hline
15 & $-$ & $-$ & $-$ & $-$ & $-$ & $-$ & $-$ & $-$ & $-$ & $-$ & $-$ & $-$ & \textbf{11.50} & $-$ & \textbf{11.65} \\
\hline
16 & $-$ & $-$ & $-$ & $-$ & $-$ & $-$ & $-$ & $-$ & $-$ & $-$ & $-$ & $-$ & \textbf{11.50} & $-$ & \textbf{11.25} \\
\hline
17-22 & $-$ & $-$ & $-$ & $-$ & $-$ & $-$ & $-$ & $-$ & $-$ & $-$ & $-$ & $-$ & \textbf{11.50} & $-$ & \textbf{10.90} \\
\hline
23-28 & $-$ & $-$ & $-$ & $-$ & $-$ & $-$ & $-$ & $-$ & $-$ & $-$ & $-$ & $-$ & $-$ & $-$ & \textbf{10.90} \\
\hline
\end{tabular}}
\label{comparison_tab}
\end{table*}

Our proposed work is compared against the SOTA approach presented in QUADRO \cite{holliday2025}. It presented two approaches for drone scheduling: a pure quantum approach and a hybrid quantum-classical approach. Their work addresses drone delivery problems, where routing is solved first, and scheduling follows as the second phase.
For comparison, we thus reproduce their routing experiments and use the resulting route completion times as input to QUACOD.
All experiments in this subsection are conducted on six instances (P-N16-K8, P-N19-K2, P-N20-K2, P-N21-K2, P-N40-K5, P-N51-K10) of the Augerat dataset \cite{augerat1995computational} adapted for the drone delivery problems \cite{uchoa2017new}.
The route completion times are measured in hours, and the recharging time is set to 1.25 hours, consistent with QUADRO \cite{holliday2025}.

Since QUACOD requires multiple iterations to converge, we measure its convergence process across iterations, as illustrated in Fig.~\ref{convergence_plots_on_QUADRO_dataset}.
Due to the inherent instability of the VQE algorithm, suboptimal solutions are produced at certain iterations.
Nevertheless, QUACOD consistently converges to the optimal solution in the long run.

Finally, we compare QUACOD with the hybrid and pure quantum approaches in QUADRO, as shown in Table~\ref{comparison_tab}.
QUACOD not only outperforms both QUADRO approaches but also finds solutions in all cases, whereas QUADRO fails to do so when the number of drones and routes is large. These results empirically demonstrate the solution quality and scalability of QUACOD.

\subsection{Benchmark on Large-Scale Instances}

To demonstrate the scalability of QUACOD, we conduct experiments on larger-scale instances from the Benchmark Problems for Parallel Machine Scheduling dataset \cite{wang2019effective}.
Originally designed for parallel machine scheduling, we adapt it to the drone scheduling problem by treating jobs as routes and machines as drones, retaining only the first machine's completion time for each job as the route completion time, measured in minutes.
The recharging time remains at 75 minutes (1.25 hours), as in Section~\ref{QUADRO}.
The dataset contains 21 configurations of nc instances, and for each configuration, we select the first instance for evaluation, as listed in Table~\ref{large_scale_instances}.

\begin{table}[t]
    \centering
    \caption{Results of selected instances from \cite{wang2019effective} adapted for the drone scheduling problem, where the fourth column shows the time in minutes until all drones have completed all routes.}
    \label{large_scale_instances}
    \renewcommand{\arraystretch}{1.3}
    \begin{tabular}{|c|c|c|c|}
        \hline
        Number of routes & Number of drones & Instance & QUACOD \\
        \hline
        \multirow{4}{*}{200} & 5 & mj5-200nc1 & 5135 \\
        \cline{2-4}
        & 10 & mj10-200nc1 & 2476 \\
        \cline{2-4}
        & 20 & mj20-200nc1 & 1215 \\
        \cline{2-4}
        & 50 & mj50-200nc1 & 471 \\
        \hline
        \multirow{4}{*}{250} & 5 & mj5-250nc1 & 6485 \\
        \cline{2-4}
        & 10 & mj10-250nc1 & 3129 \\
        \cline{2-4}
        & 20 & mj20-250nc1 & 1567 \\
        \cline{2-4}
        & 50 & mj50-250nc1 & 612 \\
        \hline
        \multirow{4}{*}{300} & 5 & mj5-300nc1 & 7652 \\
        \cline{2-4}
        & 10 & mj10-300nc1 & 3839 \\
        \cline{2-4}
        & 20 & mj20-300nc1 & 1912 \\
        \cline{2-4}
        & 50 & mj50-300nc1 & 733 \\
        \hline
        \multirow{4}{*}{500} & 5 & mj5-500nc1 & 12824 \\
        \cline{2-4}
        & 10 & mj10-500nc1 & 6486 \\
        \cline{2-4}
        & 20 & mj20-500nc1 & 3217 \\
        \cline{2-4}
        & 50 & mj50-500nc1 & 1253 \\
        \hline
        \multirow{4}{*}{1000} & 5 & mj5-1000nc1 & 25786 \\
        \cline{2-4}
        & 10 & mj10-1000nc1 & 12817 \\
        \cline{2-4}
        & 20 & mj20-1000nc1 & 6470 \\
        \cline{2-4}
        & 50 & mj50-1000nc1 & $-$ \\
        \hline
        3000 & 100 & mj100-3000nc1 & $-$ \\
        \hline
    \end{tabular}
\end{table}

For each instance, QUACOD is run for 400 iterations.
The results are presented in Table~\ref{large_scale_instances}.
QUACOD successfully runs on all instances except the last two, which fail due to out-of-memory errors.
Since the number of terms (both linear and quadratic) in Eqn.~\eqref{final_obj_func} is $O(n^2q^2)$, storing the corresponding variables for the last two instances requires billions of entries, which can easily exceed the capacity of a 16GB RAM system.
Despite this limitation, QUACOD solves most large-scale instances with a limited number of qubits, demonstrating its scalability.

\section{Conclusion}\label{conclusion}

In this work, QUACOD has demonstrated its potential in solving large-scale drone scheduling problems. Specifically, it has outperformed the SOTA method, QUADRO, in terms of maximum drone completion time across the same problem instances.
Moreover, it has found solutions for problems with up to 5 times as many drones and 35 times as many routes as QUADRO.
All results generated by QUACOD have been simulated using a limited number of qubits (20 qubits) and a hardware-efficient VQE ansatz, demonstrating its potential for deployment on larger real quantum hardware to solve practical large-scale problems.

Nevertheless, QUACOD also has its limitations. First, this work is evaluated only on classical hardware through simulation, not on real quantum devices. Therefore, it does not yet account for adverse effects arising during implementation in a quantum environment, such as decoherence, cross-talk, leakage, noise, and errors. Second, although the coordinate descent component performs well in practice, it lacks formal theoretical proofs of convergence and convergence rate. In addition, while VQE shows good empirical performance, no known theoretical proofs support its advantage in optimization problems. Third, the analysis of QUACOD could be extended further, for instance by examining the bottleneck of the classical component in large-scale cases or by comparing against SOTA purely classical algorithms. Lastly, the drone scheduling assumptions adopted in this work are relatively simple and do not fully capture the complexity of real-world logistics systems that integrate multiple types of vehicles, of which drones are only one. We plan to address these aspects in future work.

\section*{Acknowledgment}

This work is partly supported by MonArk NSF Quantum Foundry (DMR-1906383), NSF Quantum Award (2444042), and NSF EAGER (2602772). Any opinions, findings, and conclusions or recommendations expressed in this material are those of the author(s) and do not necessarily reflect the views of the National Science Foundation.

\bibliographystyle{IEEEtran}
\bibliography{references}

\end{document}